\documentclass[10pt,a4paper,twocolumn,english,aps,showpacs,floatfix,groupedaddress,superscriptaddress,prb]{revtex4}

\usepackage{graphicx} 
\usepackage{epsfig} 
\usepackage[english]{babel}
\usepackage{amsmath} 
\usepackage{amssymb} 
\usepackage{amstext}
\usepackage{amsfonts}
\usepackage{longtable} 
\usepackage{dcolumn} 
\usepackage{bm}
\usepackage{bbm}
\usepackage{dsfont}

\usepackage{subfigure}
\usepackage{tabularx}
\usepackage{color}
\usepackage{braket}

\begin{document}

\title{Dynamics and decoherence in the central spin model in the low-field
limit}

\author{Daniel Stanek} 
\email{daniel.stanek@tu-dortmund.de}
\affiliation{Lehrstuhl f\"{u}r Theoretische Physik I, Technische Universit\"at
  Dortmund, 44221 Dortmund, Germany} 
\author{Carsten Raas}
\email{carsten.raas@tu-dortmund.de}
\affiliation{Lehrstuhl f\"{u}r Theoretische
  Physik I, Technische Universit\"at Dortmund, 44221
  Dortmund, Germany} 
\author{G\"otz S. Uhrig}
\email{goetz.uhrig@tu-dortmund.de}
\affiliation{Lehrstuhl f\"{u}r Theoretische
  Physik I, Technische Universit\"at Dortmund, 44221
  Dortmund, Germany}

\date{\rm\today}

\begin{abstract}
We present a combination of analytic calculations and a powerful numerical
method for large spin baths in the low-field limit. The hyperfine interaction between 
the central spin and the bath is fully captured by the density matrix renormalization group. The adoption of the
density matrix renormalization group for the central spin model is presented and a proper method
for calculating the real-time evolution at infinite temperature is identified. In addition, we
study to which extent a semiclassical model, where a quantum spin-1/2 interacts with a bath
of classical Gaussian fluctuations, can capture the physics of the central spin model. The model is treated by average Hamiltonian
theory and by numerical simulation.
\end{abstract}


\maketitle

\section{Introduction \& motivation}

During the past two decades, the field of quantum information
processing\cite{Nielse2010,Stolze2008} has undergone a massive development.
One of the biggest remaining challenges is finding a way to diminish the effects of decoherence. Only if the
coherence time of the quantum bit is sufficiently long, 
information can be stored and computations can be performed. Hence, it is
essential to understand the underlying
mechanisms for the decoherence in great detail. 

A promising candidate for the realization of a quantum bit is an electron
confined in a quantum dot.\cite{Schlie2003}
 Examples for other candidates are ionic traps\cite{Leibfr2003} and nitrogen and
phosphorus atoms embedded in C$_{60}$-fullerenes.\cite{Morton2006} 
Possible candidates are also found in solids, for example single nitrogen-vacancy centers in diamond.\cite{Jelezk2004} 

In this work, we focus on a localized electron in a quantum dot. As many
semiconducting materials contain sizeable fractions of isotopes with substantial nuclear magnetic moments, the dominating
mechanism for the decoherence in the quantum dot
is the hyperfine interaction between the electron spin and the surrounding
nuclear spins.\cite{Schlie2003} Other possible mechanism could be the spin-orbit
coupling with impurity scattering
and the spin-orbit coupling with electron-phonon interaction. 
{But it has been shown that the relaxation of the electron spin due to spin-orbit coupling is strongly suppressed for localized electrons.\cite{Khaets2000,Khaets2001} 
Furthermore, a consecutive study of the dephasing of the electron spin under spin-orbit coupling revealed unrealistic dephasing times $T_2=2T_1$.\cite{Golova2004} Hence, the contact hyperfine interaction is the essential
mechanism for the decoherence of the electron spin in the quantum dot.

The dipolar interaction between
the nuclear spins takes place on a time scale which is about 1-2 orders of
magnitude larger than the time scale of the hyperfine action. Thus, it is negligible for the time scales
discussed in this paper.

The hyperfine interaction of a localized electron in a quantum dot is well
described by the central spin or Gaudin model\cite{Gaudin1976,Gaudin1983}
\begin{align}
\begin{split}
 H&=\vec{S}_0\sum\limits_{i=1}^N J_i\vec{S}_i \\ \label{eq:CSM}
  &=\vec{S}_0\cdot\vec{A},
\end{split}
\end{align}
where $N$ is the number of bath spins.
Without loss of generality, we restrict the discussion in the present work to
spin-1/2. The coupling constant $J_i$ is proportional to the probability $|\Psi(r_i)|^2$ that the electron is present at the site $r_i$ of the nuclear spin $i$.
The time scale for the fast dynamics is set by $1/J_\mathrm{q}$ where $J_\mathrm{q}^2$ is given by the quadratic sum of all couplings\cite{Merkul2002} 
\begin{align}
\sum\limits_{i=1}^N J_i^2 &=:J_\mathrm{q}^2\label{eq:Jq}
\end{align} 
which results from the distribution of the couplings in the quantum dot.
As the distance between the randomly located bath spins and the central spin $\vec{S}_0$ varies, the $\{J_i\}$ are
distributed randomly. An example for a distribution in a spherical
quantum dot is discussed in Ref.~\onlinecite{Schlie2003}. The contact hyperfine interaction is
regarded as isotropic, since we assume that the electron is in its orbital
ground state. 

In the present work, we aim at a proof-of-principle investigation of the central spin model. Thus, we
do not focus on distributions appropriate for particular experimental setups. Rather, we use a generic uniform distribution $J_i\in[0,J_\mathrm{c}]$, 
where the cutoff $J_\mathrm{c}$ is determined by the total energy $J_\mathrm{q}$. In our calculations, the randomness in the interaction constants 
is avoided by picking equidistant couplings from the box $[0,J_\mathrm{c}]$
\begin{align}
 J_i&=\sqrt{\frac{6N}{2N^2+3N+1}}\frac{N+1-i}{N} J_\mathrm{q}, \ \ i\in\{1,\ldots,N\}.\label{eq:Ji}
\end{align}
The choice~\eqref{eq:Ji} fulfills the normalization constraint~\eqref{eq:Jq}.
Hence, the relaxation always takes place on the same time scale, independent of the
actual bath size. In the following, the time $t$ is always denoted in units of $J_\mathrm{q}^{-1}$. 
{All numerical calculations were carried out for $J_\mathrm{q}=1$.}

The physics of the central spin model~\eqref{eq:CSM} has been studied with a large
variety of different methods. For example, the Bethe ansatz has been used intensively to study the
model.\cite{Gaudin1976,Gaudin1983,Bortz2007a,Bortz2007b,Bortz2010,Bortz2010a}
But the solutions are usually restricted to highly polarized states where only a
small number of bath spins is flipped. Recently, a method combining the
algebraic Bethe ansatz and Monte Carlo sampling 
has been introduced.\cite{Fariba2013,Fariba2013a} This approach is not
restricted to states with a strong polarization and it is used for calculating the
real-time evolution up to long times $\approx$ 100-1000 $[J_\mathrm{q}^{-1}]$. So far, up to $N=48$ bath spins have
been studied which
exceeds the number of spins accessible by exact diagonalization techniques.
\cite{Schlie2003,Cywins2010}

Other approaches are based on cluster expansion techniques which are usually
restricted to the strong field limit \cite{Witzel2005,Witzel2006,Maze2008,Yang2008a,Yang2009}
where spin flips between bath and central spin are neglected. First attempts have been made to include 
the flip-flop terms between central spin and bath on the level of a 1-cluster contribution.\cite{Witzel2012} A non-Markovian
master-equation formalism\cite{Coish2004,Fische2007,Ferrar2008} also yields best
results for
large magnetic fields. Recent results for the low-field limit\cite{Barnes2012}
are restricted to short time scales. 
  
In the present work, we use the time-dependent density matrix renormalization
group (tDMRG) to calculate the real-time evolution in the central spin model.
This numerical approach fully captures the hyperfine interaction for systems
consisting of up to $N\approx 1000$ nuclear spins.
Arbitrary fields can be applied to all spins, no restriction to a defined regime
has to be made. In the present work, we focus on the isotropic model without any field. This low-field limit of the central spin model~\eqref{eq:CSM}
is technically most demanding because no simplifications can be made to the model. For instance, spin-flips between central spin and bath, which are often
excluded in cluster expansions, have a sizeable contribution in the absence of a magnetic field.

This paper is structured as follows. First, we describe how the DMRG algorithm is implemented for the central spin
model \eqref{eq:CSM} and discuss an efficient way for calculating the real-time
evolution
of observables. Second, a semiclassical approximation of the central spin
model is introduced which is analyzed in the framework of average Hamiltonian
theory (AHT). Finally, we compare the result of the AHT with a numerical simulation of the semiclassical model and the tDMRG
results for the central spin model. If the number of bath spins is large enough ($N\approx 100-1000$), the
semiclassical model reproduces indeed the physics of the central spin model.

  \section{Density matrix renormalization group}
    
    Since its introduction by White in 1992,\cite{White1992,White1993} the
density matrix renormalization group has become one of the leading
numerical methods for studying 
    the physics of one dimensional quantum systems. In the last two decades, a
large number of modifications and extensions of the original DMRG algorithm has
been developed. For a review, see for instance Refs. \onlinecite{Scholl2005} and \onlinecite{Scholl2011}.
    In this section, we present how the algorithm can be implemented for a
cluster of spins linked by a central spin as it occurs in the central spin model. Furthermore, an
appropriate choice for the initial state and various methods for calculating the
real-time evolution are discussed.

  \subsection{Clusters \& DMRG}

  \begin{figure}[tb]
   \includegraphics[height=2cm]{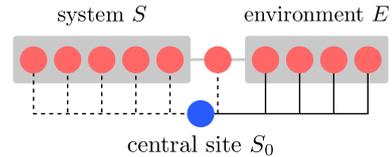}
   \caption{(Color online) Exemplary realization of the DMRG algorithm for the central spin
model with 10 bath spins. For clarity, we refrain from showing the central
spin as a part of the environment. The dashed and solid lines mark the
interaction between the central spin and the system and the environment block,
respectively.} \label{fig:DMRG}
  \end{figure}

    The implementation of DMRG algorithms for one dimensional structures like
chains is well established. But how can a cluster in
the central spin model \eqref{eq:CSM} be studied?
    For the beginning, let us neglect the central spin.
The bath spins in \eqref{eq:CSM} form a non-interacting chain of spins, which
can be split into the well-known DMRG setup of system and environment block. The finite-size
algorithm\cite{White1992,White1993} can be used to sweep through the entire
chain. In contrast to the original finite-size algorithm, we use a one-site
algorithm\cite{White2005} where only a single site is moved from the environment block to the system block in each step of a sweep.

    Next, we consider the central spin.
    The key point for the success of the algorithm lies in its exact treatment
so that the DMRG optimizes only the representation of the bath sites. By keeping the
central spin separate from the bath spins, its operators and thus the
interaction with the surrounding bath spins is treated exactly. This procedure corresponds to a representation
with three basis sets: system, environment, and a single site (the central spin). To circumvent the bookkeeping
for the single site state and to make the algorithm closer to standard DMRG techniques, it is convenient 
to integrate the single site into the environment. Then two spins have
to be added in each step of the algorithm: A bath spin to the system block and the central spin to the environment
block. For convenience, old system blocks should be reused as environment
blocks. The truncation with respect to the reduced density matrix only affects
the basis of the system block. Thus, the basis of the central spin is never reduced
and always stored exactly. At this point, it also becomes clear why we choose a
one-site algorithm. In each step, one already has to treat two sites exactly, namely one bath site and the central site. 
{The separate treatment of the central site in the framework of DMRG was already addressed before.\cite{Friedr2006}
In this approach, the implementation was based on the standard two-site DMRG algorithm so that three sites have to be treated exactly.
The additional exact site implies a doubling of the Hilbert space which results in a noticeable decrease of the performance of the algorithm. 
Hence, the employment of the one-site DMRG for the central spin model is favorable. Thereby, we are able to study systems of up to $\approx 1000$ bath spins at infinite
temperature. In the previous work for the two-site DMRG, the results were limited to 99 bath spins at most depending on the studied initial state.~\cite{Friedr2006}}   

    An exemplary setup for 10 bath spins is sketched in Fig.~\ref{fig:DMRG}. For
clarity, the central spin is depicted separate from the environment. The solid
lines indicate the interaction between the central spin and all bath spins in
the environment block. The interaction terms are added when the central spin is integrated into the
environment block.
    The dashed lines label the interaction between central site and system
block, which is calculated only when the action of the super block Hamiltonian
on the superblock wave vector is required. Note that the 
system block Hamiltonian is always zero. 

    The realization of the DMRG for clusters requires only a small amount of
changes to an existing DMRG code, which mainly affect the way how a bath or
the central spin is added to a block. For the sweeps, the transformation of the
wave vector of the super block has to modified, but this is also
straightforward. 

Furthermore, an interaction between bath spins to yield a wheel topology with interaction between adjacent spins would be possible as well. But the discussion
of such interactions lies beyond the present work. They can arise, e.g., from dipolar couplings.

 \subsection{Purification}

  \begin{figure}[tb]
   \includegraphics[width=\columnwidth]{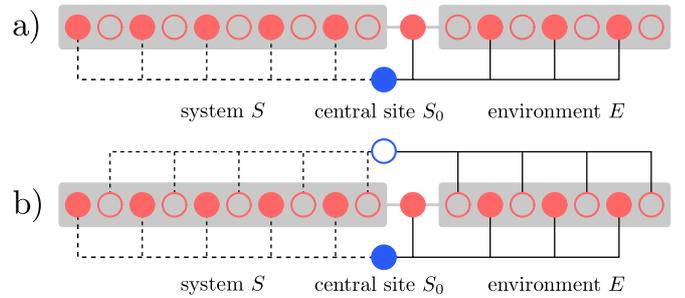}
\caption{(Color online) Exemplary realization of the DMRG algorithm for the central spin model
for a purified system. The real sites are represented by solid dots, the
auxiliary ones by circles. The dashed and solid lines mark the interaction
between the central site and the system and the environment block,
respectively. In the lower setup, the central site is purified as well.} \label{fig:DMRG_purified}
  \end{figure}

    Our goal is to calculate the time evolution of observables and correlation
functions at room temperature. For the nuclear spins in the bath, this
corresponds to infinite temperature because the distance of their energy levels is very small\cite{Merkul2002,Schlie2003} 
compared to the thermal energy at $T\approx 300$~K.
Then, all initial states are equiprobable
and the expectation value of an observable reduces to its mean value with
respect to all possible initial states. As the size of the Hilbert space grows
exponentially with the number of spins, a direct evaluation of the corresponding traces is out of question.
We therefore have to develop a strategy to calculate the traces as
accurately as possible. 

    One possibility is to calculate the expectation values for only a small
number of randomly chosen initial states. Then, the final result is given as the
average over all of these random initial states. 
    This method has proven to perform very well in the framework of exact
diagonalization. However, for a fast and reliable convergence it is
required to use arbitrary superpositions of basis states as initial state.\cite{Silver1994} Such states do not have a fixed quantum number.

In contrast, it is advisable to use states with a well-defined quantum number for 
the DMRG calculation. Thereby, the conservation
of the total magnetization can be exploited which permits a significant increase of
the performance of the algorithm. Of course, the sampling can be done with
    simple basis states as initial states. But a few hundred of them will be
required to yield reliable results for the traces, in contrast to a handful of arbitrarily superpositioned states in an exact diagonalization. Furthermore, the
resulting expectation value always suffers from the sampling of a finite number of initial states. For instance, systematic
extrapolations in the system size are hardly possible because of the
insufficient precision of this method.

    A method which allows us to calculate the expectation values at $T=\infty$
exactly is known in the literature under the key word
\textit{purification}.\cite{Buehle2000,Karras2012} By introducing an auxiliary
spin to each real spin, the size of the system is doubled. In the following, only the bath spins have an
auxiliary spin.  We show that in our case the purification of the central spin is optional.
A more detailed discussion concerning an auxiliary central spin follows below. 

    At $t=0$, each real spin is prepared in a singlet state with its
corresponding auxiliary spin. The initial state of the bath is given by
    \begin{align}
     \ket{S}&:=\bigotimes\limits_{i=1}^N
\frac{1}{\sqrt{2}}\left(\ket{\uparrow_\text{r}\downarrow_\text{a}}-\ket{\downarrow_\text{r}\uparrow_\text{a}
}\right)_i,
    \end{align}
    where $\mathrm{r}$ stands for the real spin and $\mathrm{a}$ for the auxiliary spin,
respectively. The total initial state 
    \begin{align}
     \ket{\Psi_\alpha(0)}&=\ket{S}\otimes\ket{\alpha}_0\label{eq:ini_pur}, \ \ \ \alpha\in\{\uparrow,\downarrow\}
    \end{align}
    is the product state of $\ket{S}$ and the state of the central spin which can be $\uparrow$ or $\downarrow$. All
operators are restricted to real spins
    \begin{align}
     \widehat{O}\longrightarrow \widehat{O}_\text{r}\otimes\mathds{1}_\mathrm{a}
    \end{align}
    so that the artificial doubling of the bath does not affect the physics of the model. 

The key observation is that the trace of an observable in the space of the real
spins is reduced 
    to a simple expectation value in the extended Hilbert space\cite{Buehle2000}
    \begin{align}
     \left.\mathrm{Tr}\,\widehat{O}\left(t\right)\right|_\text{r}&=\left.\frac{\bra{\Psi_\uparrow\left(t\right)}
\widehat{O}\ket{\Psi_\uparrow\left(t\right)}+\bra{\Psi_\downarrow\left(t\right)}
\widehat{O}\ket{\Psi_\downarrow\left(t\right)}}{2}\right|_{\text{r}\otimes \text{a}},
    \end{align}
    which is taken with respect to the purified initial state from Eq.~\eqref{eq:ini_pur}. We underline that this expression allows for an exact
calculation of the trace with only one or two states, depending on the symmetry of the model. The initial state \eqref{eq:ini_pur} has
a well defined quantum number which makes it particularly suitable for the DMRG.	

    In the isotropic model, the system is symmetric under a spin flip of the
central spin. Thus, it is sufficient to consider only one initial state where
the central spin at $t=0$ points either up or down. If the spin flip symmetry is broken, one has to run
two independent calculations for both states of the central spin. The total
trace is then simply given as the mean value of both results.
    Compared to a calculation with a purified central spin, this procedure leads
to a smaller truncation error and shorter run-time.

    A generic DMRG setup for a purified system is sketched in Fig.~\ref{fig:DMRG_purified}, panel a).
 The auxiliary spins (circles) are integrated into the
bath. The real spins (dots) always have odd indices, while the auxiliary ones
are always indexed by even numbers. For the sake of completeness, the setup including a purified central
spin is shown as well, panel b). 

   Starting with the initial state \eqref{eq:ini_pur}, finite temperatures are
reachable by calculating the time evolution in imaginary time.\cite{Barthe2009} For the physics under study in 
the present article we do not need to pursue this option

In the following sections, we investigate three
different methods for calculating the real-time evolution using DMRG.

  \subsection{Trotter-Suzuki decomposition}
  
  The adaptive method based on a Trotter-Suzuki (TS) decomposition of the time-evolution operator
was among the first methods for calculating the real-time evolution with
DMRG.\cite{White2004,Daley2004} The Hamiltonian $H$ has to be decomposed into
local parts
  \begin{align}
   H&=\sum\limits_{i=1}^Nh_i\,,\label{eq:H_dec}
  \end{align}
  where the bond $i$ contains the interaction between bath spin $i$ and the
central spin. Then, the Trotter-Suzuki decomposition is used to split the time
evolution operator 
  \begin{align}
   U:=U\left(t,t+\Delta t\right)&=e^{-\mathrm{i}H\Delta t}
  \end{align}
  into local parts. In 2nd order, one obtains\cite{White2004} 
  \begin{align}
    \begin{split}
   U&=e^{-\mathrm{i}h_1\frac{\Delta t}{2}} e^{-\mathrm{i}h_2\frac{\Delta t}{2}}\ldots
e^{-\mathrm{i}h_{N-1}\frac{\Delta t}{2}} e^{-\mathrm{i}h_N\Delta t} \\ &\quad\times
e^{-\mathrm{i}h_{N-1}\frac{\Delta t}{2}}\ldots e^{-\mathrm{i}h_2\frac{\Delta t}{2}}
e^{-\mathrm{i}h_1\frac{\Delta t}{2}}+\mathcal{O}\left(\Delta t^3\right).\label{eq:ST_2nd}
    \end{split}
\end{align}
  
The decomposition is used to apply all local time-evolution operators
successively to their corresponding configuration of bath and central spin while
sweeping through the superblock. During a complete backwards and forwards sweep, each
bath spin is addressed twice and the local time-evolution
  operator can be applied to the individual configuration of bath and central
spin without any additional error. The action of the local $U_i$ is calculated in each step of a sweep during the transformation of the wave vector to the new
basis of the system and environment block. This transformation of the wave vector is compulsory in the finite-size algorithm and only a few
changes have to be made to an existing code to implement the time
evolution. The local parts of $U$ in Eq.~\eqref{eq:ST_2nd} are small
matrices which only act on the local configuration. They are either known
exactly or they can be computed with small numerical effort. 

  Hence, the real-time evolution from $t$ to $t+\Delta t$ is performed within
one complete backwards and forwards sweep through the superblock. The Trotter-Suzuki decomposition
\eqref{eq:ST_2nd} introduces an additional error to the truncation error. 
  For the 2nd order, the error is $\sim\Delta t^3$. As the decomposition is
applied $1/\Delta t$ times, the total error sums up to $\mathcal{O}(\Delta t^2)$
in 2nd order. 

Higher orders of the Trotter-Suzuki decomposition reduce this
error. The 4th order is derived in Appendix~\ref{app:ST}. Thereby, one step in
the real-time evolution is performed with three backwards and forwards sweeps. The
Trotter-Suzuki error is of $ \mathcal{O}(\Delta t^5)$. Thus, the error due to the
decomposition is decreased by two orders of 
  magnitude by increasing the run-time by a factor of three. However, it is not
always necessary to use higher orders of the decomposition because the Trotter-Suzuki
error usually dominates the total error only for small times. In contrast, the truncation error accumulates with increasing time.
Thus, it is generically the truncation error which limits the accuracy in tDMRG. A detailed discussion of the
Trotter-Suzuki error can be found in Ref.~\onlinecite{Gobert2005}.
  
  \subsection{Krylov vectors}

    The presented Trotter-Suzuki approach takes advantage of the features of the
finite-size algorithm but it is restricted to Hamiltonians which can be
decomposed according to Eq.~\eqref{eq:H_dec}. Furthermore, the Trotter-Suzuki error
occurs in addition to the truncation error. 

A different ansatz is based on the direct application of the time-evolution
operator $U$ to the wave vector so that a decomposition is not required. As for the
Hamiltonian $H$ of the super block, there is no complete representation of the
time-evolution operator $U$. Thus, $U$ has to be expanded in a well-defined basis so
that its action on the wave vector can be evaluated directly. The idea behind
this approach is that first the basis is optimized in the time-interval $t$ and $t+\Delta t$.
This is done by targeting several states $\Psi(t_i)$ for $t_i\in[t,t+\Delta t]$. Usually, a few half
sweeps are sufficient to optimize the basis. Afterwards, the time evolution of
the wave vector from $t$ to $t+\Delta t$ is calculated. To reduce the
integration error, it is possible to use smaller time steps for the evolution
than for the calculation of the target states. In a first
approach,\cite{Feigui2005} a Runge-Kutta integration was used to calculate the
target states and the real-time evolution. 
    But Feiguin and White also suggested to employ other methods, e.g., a Lanczos
tridiagonalization of the Hamiltonian. In contrast to the Runge-Kutte integration, this approach
preserves unitarity. However, we have to keep in mind that unitarity is always violated
by the DMRG truncation of the Hilbert space. 

    In our study, we use Krylov vectors\cite{Hochbr1997,Hochbr1999,Noack2005,Manman2005}
for the calculation of the target states as well as for the real-time evolution.
Therefore, the wave vector at $t+\Delta t$ is expanded in the basis of the
Krylov subspace
    \begin{align}    
\left\{\ket{\Psi\left(t\right)},H\ket{\Psi\left(t\right)},H^2\ket{
\Psi\left(t\right)},\ldots,H^n\ket{\Psi\left(t\right)}\right\}.
    \end{align}
    The subspace is spanned by the so called Krylov vectors  $\ket{v_n}$ which are
orthogonalized for the expansion of $\ket{\Psi(t+\Delta t)}$ via the
well-known Lanczos tridiagonalization. The recursion relation is given by
    \begin{align}
     \ket{v_{n+1}}&=H\ket{v_n} - \alpha_n\ket{v_n} - \beta_n^2\ket{v_{n-1}}, 
    \end{align}
    where the previous two vectors and the coefficients
   \begin{subequations}
     \begin{align}
      \alpha_n&=\ \frac{\bra{v_n}H\ket{v_n}}{\braket{v_n|v_n}} \\
      \beta_n^2&= \frac{\braket{v_n|v_n}}{\braket{v_{n-1}|v_{n-1}}}
     \end{align}
    \end{subequations} 
  enter. We consider a $m$-dimensional Krylov subspace, where $m$ is a very
small number compared to the dimension of the Hilbert space. In the Krylov
subspace, the Hamiltonian
  \begin{align}
   T_m&=V_m^\top H V_m
  \end{align}
  is tridiagonal and can be diagonalized easily. The matrix $V_m$ contains all
$m$ Krylov vectors of the subspace as columns. 
  Now, we approximate the time-evolution operator in the Krylov
subspace
  \begin{align}
   \ket{\Psi\left(t+\Delta t\right)}&\approx{V}_m e^{-\mathrm{i} T_m\Delta t}
{V}_m^\top\ket{\Psi\left(t\right)}. \label{eq:Krylov_approx}
  \end{align}
  The application of $V_m^\top$ onto the wave vector $\ket{\Psi(t)}=\ket{v_0}$ reduces to $V_m^\top\ket{\Psi(t)}=\ket{\Psi(t)}$ because all Krylov vectors are orthogonal to each other. In addition, the
tridiagonal Hamiltonian $T_m$ is diagonalized by an orthogonal transformation $O_m$
  \begin{align}
   e^{-\mathrm{i}T_m\Delta t}&=O_m^{\phantom{\top}} e^{-\mathrm{i}D_m\Delta t} O_m^\top,
  \end{align}
  where $D_m$ is diagonal. By inserting the latter expression in
\eqref{eq:Krylov_approx}, the expression for the real-time evolution from
$t\rightarrow t+\Delta t$ is given as
  \begin{align}
   \ket{\Psi\left(t\right)}&=\sum\limits_{i=1}^m a_i \ket{v_i}
  \end{align}
  with the coefficients
  \begin{align}
   a_i&=\sum\limits_{j=1}^m \braket{v_i|n_j}e^{-\mathrm{i}\Delta
t\lambda_j}\braket{n_j|\Psi\left(t\right)}.
  \end{align}
  The eigenvectors $\{\ket{n_j}\}$ and eigenvalues $\{\lambda_j\}$ result from the
diagonalization of $T_m$, which has to be calculated in addition to the
orthogonalized Krylov vectors $\{\ket{v_j}\}$.
  
  The coefficients $a_i$ decay extremely fast with increasing order $m$. Thus,
the dimension $m$ of the Krylov subspace can be kept very small. The modulus of the coefficients can
be used as convergence criterion. In practice, we omit contributions with $|a_i|<10^{-10}$ and only a handful of Krylov vectors is required.
  
  For the target states, we stick to the choice of four states made by Feiguin
and White:\cite{Feigui2005}
\begin{align}  \ket{\Psi (t)},\ket{\Psi (t+\Delta t/3)},\ket{\Psi(t+2\Delta t/3)}, \ket{\Psi (t+\Delta t)}.
\end{align}

The use of the Lanczos algorithm in the Krylov approach results in
longer run-times compared to the previously introduced Trotter-Suzuki decomposition, as the action of the Hamiltonian on the wave vector has to be
calculated multiple times in every step.

  \subsection{Chebychev expansion}
    The Chebychev expansion\cite{TalEze1984} is a widely known approach for
calculating the time evolution in the field of exact diagonalization. Recently,
it has also been implemented in the framework of matrix product states.
\cite{Holzne2011}
    Here, we illustrate how the Chebychev expansion is used to calculate the
real-time evolution of autocorrelation functions with DMRG. 
	
    As initial state, we consider an entirely purified initial state $\ket{0}$
as sketched in the lower panel in Fig.~\ref{fig:DMRG_purified}. The auxiliary central spin gives us
the opportunity to introduce an artificial time evolution for all auxiliary spins. An
arbitrary unitary transformation can be used, since the physics of the real system is
not affected. We follow the proposal of Karrasch \textit{et
al}.\cite{Karras2012} They studied a purified Heisenberg chain and used the
same Hamiltonian for the auxiliary spins but calculated their real-time
evolution
    backwards in time. In the Heisenberg chain, this leads to a slower growth of
the entanglement and thus to a slower increase of the truncation error.

    We consider an autocorrelation function of the central spin in the
Heisenberg picture,  e.g., in $z$-direction,
    \begin{align}    
S\left(t\right)&=\braket{0|U^\dagger\left(t\right)S^z_0U\left(t\right)S^z_0|0},
\label{eq:Cheby_S}
    \end{align}
 where $U(t)=e^{-\mathrm{i}tH}$  with $H=H_\mathrm{r} - H_\mathrm{a}$ acts on the
real \textit{and} auxiliary spins. As the real spins evolve forward and the
auxialiry spins evolve backward in time, the action of $U^{(\dagger)}$ on
$\ket{0}$ leaves $\ket{0}$ unchanged. This observation sets the basis for the reduced growth of entanglement in
chain topologies. For a more detailed discussion, see Appendix~\ref{app:Pur}. Thus, the autocorrelation function \eqref{eq:Cheby_S} reduces to
    \begin{align}
     S\left(t\right)&=\braket{0|S^z_0U\left(t\right)S^z_0|0}
\label{eq:Cheby_S2}.
    \end{align}
The Hamiltonian is rescaled with the energy bound $C=3/4\sum_{i=1}^N |J_n|$ to
ensure the validity of the expansion. The eigenvalues of the rescaled
Hamiltonian $\widetilde{H}=H/(2C)$ fulfill $-1\le \widetilde{E}_n\le 1$.     
The operator $U$ is now expanded in the basis of the Chebychev polynomials
$T_n^\mathrm{ch}(\tilde{H})$
     \begin{align}
      e^{-\mathrm{i}\tilde{H}t}&=\sum\limits_{n=0}^\infty
T_n^\mathrm{ch}\left(\frac{H_\mathrm{r}-H_\mathrm{a}}{2C}\right) b_n\left(2Ct\right). 
     \end{align}
    Note that the
time-dependence 
     resides solely in the coefficients $b_n(t)$ which read
    \begin{subequations}
     \begin{align}
      b_0\left(t\right)&=J_0\left(t\right) \\
      b_n\left(t\right)&=2\left(-\mathrm{i}\right)^n J_n\left(t\right),
     \end{align}
  \end{subequations}
    where $J_n\left(t\right)$ is the Bessel function of the first kind of order
$n$. By inserting the coefficients into the expansion, one obtains the expression 
    \begin{align}
     S\left(t\right)&=J_0\left(2Ct\right)m_0+\sum\limits_{n=1}^\infty
m_n\left(-\mathrm{i}\right)^n J_n\left(2Ct\right). \label{eq:Cheby_final}
    \end{align}
  The time-independent Chebychev polynomials are given as
    \begin{align}
     m_n&:=\braket{v_0|v_n}\label{eq:Cheby_mn}
    \end{align}
  with
  \begin{subequations}
    \begin{align}
      \ket{v_0}&=S^z_0\ket{0} \\
      \ket{v_{n+1}}&=\frac{H_\mathrm{r}-H_\mathrm{a}}{C}\ket{v_n}-\ket{v_{n-1}}.
    \end{align}
    \end{subequations}

Comparable to the Lanczos algorithm, multiple powers of the Hamiltonian have to
be calculated which is easily carried out with DMRG. As the time-dependence resides separately in
the Bessel functions, the autocorrelation function \eqref{eq:Cheby_final} can be evaluated with a 
separate code or with any computer algebra program.

For the reduced density matrix, at least the four states $\ket{v_0},
\ket{v_{n-1}}, \ket{v_n}$, and $\ket{v_{n+1}}$ have to be targeted. {The targeting of the initial state $\ket{v_0}$ is crucial because
it is required for the calculation of the coefficients $m_n$ \eqref{eq:Cheby_mn}.} 
 As in the
Krylov approach, a few half sweeps are sufficient to optimize the basis
in each step. After the optimization is completed, the Chebychev polynomial
$m_{N+1}$ is calculated and stored. Afterwards, one proceeds with the next
order. The required order
depends on the desired time~$t$. The Bessel function~$J_n(t)$ contributes
noticeable only for $t>n$. Thus, an estimate for the required order $M$ is $M\ge
2\sqrt{N}t$, where $N$ is the number of bath spins.

One has to be careful about the initial state~$\ket{v_0}$ because its
normalization is lost after multiple sweeps. Hence, we recommend to rebuild
the state in each order, which can be done during a half sweep. This reduces the
error of the Chebychev polynomials $m_n$. All others states are only stored for
three orders and a recalculation is not necessary. Concerning the weight of the
target states in the reduced density matrix, equal weights performed best
in our tests. 
A similar performance is yielded when the state $\ket{v_0}$ with weight $w_1=1/2$
dominates the reduced density matrix and all other states have the same weight
$w_i=1/6$. Different weights for the states $\ket{v_{n-1}},\ket{v_n}$ and
$\ket{v_{n+1}}$ tend to have a negative effect on the performance.  
The run-time of the DMRG code is moderate and ranks between the Trotter-Suzuki and Krylov approach.
    
  \subsection{Results}

  \begin{figure}[tb]
  \centering
   \includegraphics[width=\columnwidth]{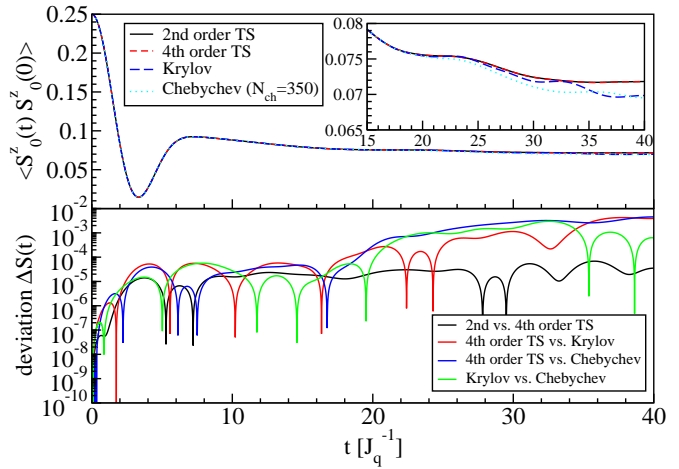}
    \caption{(Color online) Time evolution of the central spin interacting with a bath of 19 spins (upper
panel) for $1024$ states. {The inset is a magnification of the autocorrelation function for $t\ge 15$. Both the result for the
Krylov and Chebychev method deviate from the Trotter-Suzuki result for $t>20$.}
The deviation $\Delta S(t)=|S_\mathrm{a}(t)-S_\mathrm{b}(t)|$  is shown for selected pairs of methods $\mathrm{a}$ and $\mathrm{b}$ in the lower panel.} 
\label{fig:TE-comp}
 \end{figure}
  
\begin{figure}[tb]
  \centering
   \includegraphics[width=\columnwidth]{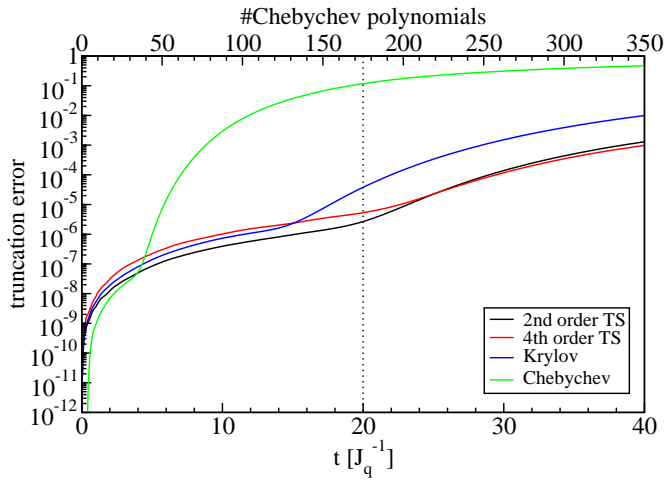}
    \caption{(Color online) Truncation error for the real-time evolution shown in Fig.~\ref{fig:TE-comp}. 
For the Trotter-Suzuki and Krylov approach, the truncation error is plotted as a function of time $t$ (lower $x$-axis),
while for the Chebychev approach it is plotted in dependence of the number of contributing polynomials (upper $x$-axis). 
The vertical dashed line marks the number of required
Chebychev polynomials up to $t=20$. While the truncation error for the Krylov
and the Trotter-Suzuki method is still acceptable at $t=20$, it is too large for
the Chebychev polynomials.}\label{fig:TruncErr}
  \end{figure}
    
     In the previous sections, we introduced three different methods for
the calculation of the real-time evolution of observables in the central spin
model. Here, we compare their results for an exemplary system of $N=19$ bath spins.
Without field, the model \eqref{eq:CSM} and all autocorrelation functions are
isotropic. The autocorrelation function $S(t)=\braket{S^z_0(t)S^z_0(0)}$ of the central spin
is plotted in Fig.~\ref{fig:TE-comp}. Its real-time evolution has been
calculated with 
    the adaptive method based on the Trotter-Suzuki decomposition in 2nd and 4th
order, time-step targeting using Krylov vectors, and the Chebychev expansion with
$350$ coefficients up to $t=40$. In addition, 
    the deviation $\Delta S(t)=|S_\mathrm{a}(t)-S_\mathrm{b}(t)|$ has been plotted in the lower panel for various pairs of
methods a and b.

    A quantitative measure for the error of the DMRG calculation is the truncation
error \begin{align}
     \text{err}&=1-\sum\limits_j w_j,
    \end{align}
    which is the sum of all truncated weights $w_j$ of the reduced density
matrix. The respective truncation error for the curves in Fig.~\ref{fig:TE-comp}
has been plotted in Fig.~\ref{fig:TruncErr}.

    We start our discussion with the autocorrelation function of the central spin in Fig.~\ref{fig:TE-comp}.
 For short times up to $t=15$-$20$, all curves agree nicely. As
shown in the inset, the result of the Chebychev expansion begins to deviate from
the other results
    at $t\approx 20$. The result of the Krylov method agrees with the
Trotter-Suzuki result up to $t\approx 25$. There is no visible distinction
between the Trotter-Suzuki result in 2nd and 4th order in the complete interval
$t\in [0,40]$. The lower panel reveals that 
    their deviation is almost constant of the order $10^{-5}$-$10^{-4}$. For the
time interval $\Delta t=0.01$ used in this calculation, this is exactly the
difference of their Trotter-Suzuki error $\propto \Delta t^2$.
    All other comparisons between the different methods show an increase of the
deviation $\Delta S(t)$ by several orders of magnitude for larger times.

    As seen in Fig.~\ref{fig:TruncErr}, there are
huge differences  in the truncation error in the different methods. The truncation error of the Chebychev
polynomials grows extremely fast. As an example, we consider
   {$N_\textrm{ch}=175$} coefficients which are required approximately to calculate the
autocorrelation function up to $t=20$, which is indicated by the dashed vertical
line in Fig.~\ref{fig:TruncErr}. At this point, the truncation error is
about 4 orders of magnitude larger than the truncation error of the
Krylov method. The truncation error of the Trotter-Suzuki approach lies even one
order of magnitude lower. The kinks in the plotted curves correspond to the
point where the number of tracked states is not sufficient anymore to capture
the growing entanglement in the model.\cite{Gobert2005}  Beyond this time, the truncation error
increases faster. This time could be used as a pessimistic upper bound for the
validity of the simulation. But from our experience, the calculation yields
correct results  even far beyond this time.

    The fast growing truncation error of the Chebychev polynomials is caused by
various mechanisms. First, the central spin has to be purified. Thus, the
total Hilbert space is twice as large as for the other methods. This leads to a
larger discarded weight per truncation. Second, all target states correspond to
different powers of the Hamiltonian. Hence, their overlap is rather small
compared to the target states used for the Krylov vector approach.
    In particular, the target state $\ket{v_0}$ is far away from the other target
states. Thus, they are difficult to represent reliably by the reduced DMRG basis.

    As a consequence, we refrain from using the Chebychev polynomials for the
central spin model. The corresponding autocorrelation function deviates too early from the other results.
Furthermore, the truncation error increases too fast. {Higher orders $N_\textrm{ch}> 350$ of the expansion do not lead to an improvement on the considered time scale $t=0$-$40\ J_\mathrm{q}^{-1}$,
 since the Bessel function $J_n(t)$ only contributes for $t>n/J_\mathrm{q}$. Additionally, the truncation error for the corresponding coefficients with $n>350$ would be $\mathcal{O}(1)$.
Hence, it is very unlikely that high orders would give a reliable contribution to the autocorrelation function.} 

{Improvement for the calculation of the Chebychev polynomials may be achieved
by a variational ansatz.~\cite{Holzne2011,Scholl2011}  
This is beyond the scope of the present work because we intended to keep the implementations close to an existing standard DMRG code. In particular, our proposal for the Chebychev expansion is closely related to the Krylov vector approach.} 
{We also} stress that the
situation in the central spin model is somehow special because of the purified
central spin. Thus, the Chebychev polynomials might be a suitable approach
for studying the real-time evolution with DMRG in other models.

    The result of the Krylov vectors is suitable up to intermediate time scales.
For large times, it deviates from both Trotter-Suzuki results as shown in Fig.~\ref{fig:TE-comp}. The run-time is about 4-5
times longer compared to the 2nd order decomposition. Furthermore, the accumulated truncation error plotted in Fig.~\ref{fig:TruncErr}
is roughly one order of magnitude larger than the one of the Trotter-Suzuki approach.

Consequently, we stick to the adaptive
approach based on the Trotter-Suzuki decomposition for all future calculations. It combines a high accuracy with
moderate run-times and a small truncation error. In addition, the access to the local time-evolution operators is easy
so that they can be modified with almost no effort to simulate time-dependent Hamiltonians. 

    The Trotter-Suzuki decomposition in 4th order is not mandatory because the total error
for large $t$ is usually dominated by the truncation error. Only for short
time scales it might be beneficial, but this is not supported by our results.   
The truncation error of the 2nd and 4th order decomposition differs only marginally, especially for long times. 

    In the introduction of the Chebychev polynomials, the backward time evolution for the
auxiliary spins in $H=H_\mathrm{r}-H_\mathrm{a}$ has already been discussed, see also Appendix~\ref{app:Pur}. As long as an auxiliary central spin is
present, this can be realized for all presented approaches. Karrasch \textit{et al.}\cite{Karras2012}
showed that the time-reversed evolution of the auxiliary spins suppresses the
growth of the entanglement. Consequently, larger time scales could be accessed
    in the real-time evolution of a Heisenberg chain. 

We implemented $H=H_\mathrm{r}-H_\mathrm{a}$ for the real-time evolution based on the Trotter-Suzuki decomposition. In contrast to previous results, 
our tests revealed no advantage compared to the scenario without purified central spin.
    The additional doubling of the Hilbert space leads to a noticeable increase of the truncation error.
Furthermore, the application of any operator to the purified initial states
creates an entanglement which then propagates in the system. Compared
to a local entanglement created in a chain,
    an entanglement created at the central spin is crucial. The central spin
directly interacts with all bath spins and the entanglement spreads almost 
immediately over the complete system. 
    In a chain, it takes much more time until the entanglement is completely
spread, cf. Appendix~\ref{app:Pur}.

  \section{Classical Gaussian fluctuations}
 
    In the previous section we discussed the fully quantum mechanical
central spin model \eqref{eq:CSM}. The heavy numerical treatment with DMRG
gives access to relatively large spin baths. But in contrast to exact
diagonalization, the accessible times are limited to $t\approx 20-40$ due to the growing
entanglement in the system.

Now we consider a semiclassical  Hamiltonian
    \begin{align}
     H&=\vec{\eta}\left(t\right)\cdot \vec{S}_0, \label{eq:HSC}
    \end{align}
    where an electron spin-1/2 interacts with a classical random field
$\vec{\eta}(t)$. The random fluctuations can be justified to be Gaussian (see below) and are fully defined by their
    mean value and their autocorrelation function
    \begin{subequations}
     \begin{align}
	\overline{\eta\left(t\right)}&=0 \\
	g_{\alpha\beta}\left(t_1-t_2\right)&=\overline{
\eta_\alpha\left(t_1\right)\eta_\beta\left(t_2\right)}\,.
      \end{align}
    \end{subequations}
    Without loss of generality, the mean value is set to zero as it only creates
a constant offset. 

The classical behavior of the bath is well supported by the properties of the quantum model~\eqref{eq:CSM}. 
{By regarding the square of an operator norm
\begin{subequations}
\begin{align}
 \mathrm{Tr}\left(A^\alpha\right)^2&=\frac{1}{4}\sum\limits_{i=1}^N J_i^2=\frac{J_\mathrm{q}^2}{4}
\end{align}
one concludes that $A^\alpha=\mathcal{O}(1)$. But for the commutator we find 
\begin{align}
  \mathrm{Tr}\left(\left[A^\alpha,A^\beta\right]\right)^2&=\frac{1}{4}\sum\limits_{i=1}^N J_i^4\propto\frac{1}{4}\sum\limits_{i=1}^N\left(\frac{J_\mathrm{q}^2}{N}\right)^2=\frac{1}{4}\frac{J_\mathrm{q}^4}{N}
\end{align}
\end{subequations}
which implies that the norm of the commutator vanishes in the limit $N\rightarrow\infty$ because $[A^\alpha,A^\beta]=\mathcal{O}(J_\mathrm{q}^2/\sqrt{N})$.}
Hence, the bath can be regarded as a classical variable for a large number of bath spins.
According to the central limit theorem, the statistics of the bath is Gaussian because it consists of a large number of independent fluctuations.

We focus on the isotropic model, where all non-diagonal
correlations vanish and all diagonal correlations are identical: $g(t)\equiv
g_{\alpha\alpha}(t)$. The semiclassical model~\eqref{eq:HSC} is derived from the
central spin model~\eqref{eq:CSM} by replacing the quantum bath $\vec{A}$ with a
classical fluctuating field $\vec{\eta}(t)$. The correlation $g_{\alpha\beta}(t)$ of the random noise
is then identified with the autocorrelation $\braket{A^\alpha(t)A^\beta(0)}$ of the bath
operators. 

    In the following, we first discuss the semiclassical model \eqref{eq:HSC}
analytically on the level of AHT. Afterwards,
the result is compared to a numerical simulation of the model. The
similarities and the differences to the central spin model are discussed in the
final step.

   \subsection{Average Hamiltonian theory}

 \begin{figure}[tb]
  \centering
   \includegraphics[width=\columnwidth]{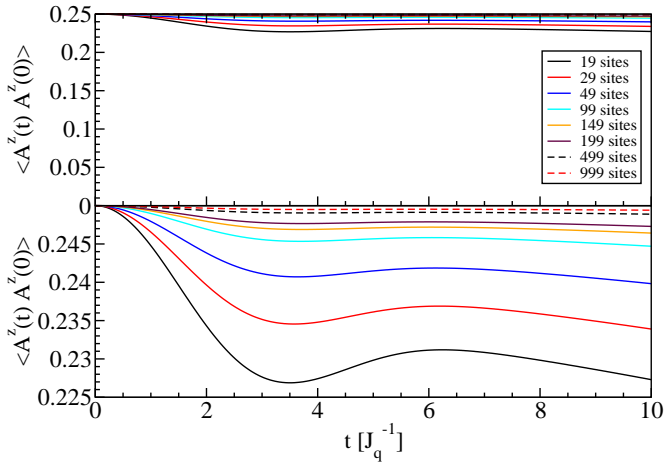}
    \caption{(Color online) Autocorrelation function of the bath spins for various system sizes
calculated with tDMRG. With increasing number of bath spins, the fluctuations
become more and more static. Note the scale of the $y$-axis in the lower panel.}
\label{fig:AzAz}
 \end{figure}	
    
      For a time-dependent Hamiltonian, the time-evolution operator is given by
	\begin{align}
	 U\left(t\right)&=\mathcal{T} \exp\left[-\mathrm{i}\int_0^t \mathrm{d}t'\,
H\left(t'\right)\right], \label{eq:U_full}
	\end{align}
    where $\mathcal{T}$ is the time-ordering operator.
     In general, there is no way to evaluate \eqref{eq:U_full} in a closed form.
We apply the Magnus expansion\cite{Magnus1954,Blanes2009} to
simplify the time-evolution operator.
      The leading order is obtained by simply neglecting the time-ordering in Eq.
\eqref{eq:U_full}. The approximation
      \begin{align}
       U\left(t\right)&\approx\exp\left[-\mathrm{i}\int_0^t \mathrm{d}t'\, H\left(t'\right)\right]\label{eq:U_1storder}
      \end{align}
      is certainly justified if the Hamiltonian is almost static which corresponds to constant autocorrelation function
of the random noise. In the central spin model, the autocorrelation
function $\braket{A^z(t)A^z(0)}$ of the bath spins is indeed almost static, as shown in Fig.~\ref{fig:AzAz}. The operator 
{$A^\alpha=\sum_{i=1}^N J_i^{\phantom{\alpha}} S^\alpha_i$} captures the local operators of all bath spins.

      By exploiting the properties of the Pauli matrices and the spherical
symmetry of the fluctuations, the time-evolution operator~\eqref{eq:U_1storder} can be written as
    \begin{align}
       U\left(t\right)&=\cos\frac{v}{2} \cdot\mathds{1} - \mathrm{i}\sin\frac{v}{2}\cdot
\sigma_{\vec{v}} \label{eq:U_sigmav}
      \end{align}       
 where $\vec{v}:=\int_0^t \mathrm{d}t'\, \vec{\eta}\left(t'\right)$. The operator 
\begin{align}
 \sigma_{\vec{v}}&=\sin\theta\cos\varphi\, \sigma_x + \sin\theta\sin\varphi\,
\sigma_y + \cos\theta\,\sigma_z
\end{align}
is the projection of the Pauli matrices onto $\vec{v}$.
The simplified time-evolution operator \eqref{eq:U_sigmav} is used to calculate
the autocorrelation function $\braket{S^z_0(t)S^z_0(0)}$ which reads
\begin{align}
 \braket{S^z_0\left(t\right)S^z_0\left(0\right)}&=\frac{1}{4}\left[\cos^2\frac{v
}{2}+\sin^2\frac{v}{2}\left(2\cos^2\theta-1\right)\right].
\end{align}

This expression still depends on the random fluctuations. Hence, its average with respect to $\theta$ and $v$ has to be calculated. Due to the spin rotational symmetry, the average over the angle $\theta$
simply yields 
$\overline{\cos^2\theta}=1/3$. The remaining contribution
\begin{align}
 \overline{\braket{S^z_0\left(t\right)S^z_0\left(0\right)}}&=\frac{1}{4}
\overline{\left(\frac{1}{3}+\frac{2}{3}\cos v\right)}\label{eq:AHT_unavg}
\end{align}
is averaged with respect to the Gaussian distribution
\begin{align}
 p\left(v\right)&=\frac{1}{\sigma\sqrt{2\pi}} e^{-\frac{v^2}{2\sigma^2}}
\end{align}
 of the fluctuations. The
final result is given as
\begin{align} 
\overline{\braket{S^z_0\left(t\right)S^z_0\left(0\right)}}&=\frac{1}{6}\left[
\left(1-\sigma^2\left(t\right)\right)e^{-\frac{\sigma^2\left(t\right)}{2}}+\frac{1}{2}\right],
\end{align}
where the time-dependence resides in the variance $\sigma^2$. It is related to
the autocorrelation function of the random noise via
\begin{align}
 \sigma^2\left(t\right)&=2\int_0^t \mathrm{d}t_1 \int_0^{t_1} \mathrm{d}t_2 \, g\left(t_1-t_2\right).
\label{eq:AHT_var}
\end{align}

Independently of the actual time-dependence of $\sigma^2(t)$, the autocorrelation
\label{eq:AHT_sigma} always converges to a plateau of $1/12$ if $\sigma^2$
increases monotonically in time. For $N\rightarrow\infty$, the autocorrelation function of the bath
in the central spin model shown in Fig.~\ref{fig:AzAz} becomes static with $g(t)=1/4$. Hence, the variance
increases quadratically in time and the autocorrelation function of the central
spin in 1st order AHT reads
\begin{align} 
\overline{\braket{S^z_0\left(t\right)S^z_0\left(0\right)}}&=\frac{1}{6}\left(e^{
-t^2/8}\left(1-\frac{t^2}{4}\right)+\frac{1}{2}\right).\label{eq:AHT_const}
\end{align}

This result is identical to the one obtained by Merkulov \textit{et
al}.\cite{Merkul2002} Their result is derived from the classical discussion of the fluctuations of a completely static spin bath.
Our AHT for the semiclassical model~\eqref{eq:HSC} is based on the Magnus
expansion and can be extended systematically to higher orders in nested commutators.\cite{Blanes2009} In addition, it includes the correlation function 
$g(t)$ of the fluctuations. The 2nd order correction of the Magnus expansion,
which leads to a renormalization of the Gaussian probability distribution, is
discussed in Appendix~\ref{app:AHT}. As the correlations in the isotropic model are
almost static, it does not lead to a recognizable improvement of the 1st order.

 \begin{figure}[tb]
  \centering
   \includegraphics[width=\columnwidth]{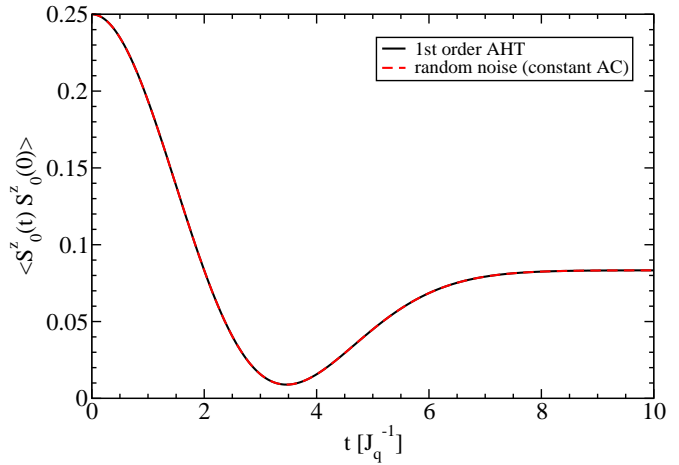}
    \caption{(Color online) Random noise simulation and 1st order AHT \eqref{eq:AHT_const}
result for the autocorrelation function of the central spin in the semiclassical
model \eqref{eq:HSC}. The correlation of the random noise is
constant.}\label{fig:AHT_const}
 \end{figure}	

Our AHT result for the semiclassical model~\eqref{eq:HSC} is verified by 
numerical simulations. The Gaussian fluctuations are sampled numerically
according to the predefined autocorrelation function $g_{\alpha\beta}(t)$. 
The integration is carried out using commutator-free exponential time
propagators (CFETs) as introduced by Alvermann \textit{et
al}.\cite{Alverm2011,Alverm2012} This reduces the integration error at the cost
of only little additional numerical effort. The result for a constant
autocorrelation function (AC) of the bath is shown in Fig.~\ref{fig:AHT_const} and reveals a perfect agreement.

\subsection{Discussion}

  The analysis of the AHT and the random noise simulation can be made for
arbitrary types of Gaussian noise. In the following, we identify the correlation function of the random noise
with the autocorrelation function of the bath spins in the central spin model. Thus, we set
\begin{align}
 g_{\alpha\beta}\left(t\right)&\stackrel{!}{=}\braket{A^\alpha\left(t\right)A^\beta\left(0\right)},
\end{align}
where the autocorrelation function of the bath $\braket{A^\alpha\left(t\right)A^\beta\left(0\right)}$ (see Fig.~\ref{fig:AzAz}) is obtained from a tDMRG calculation.
 An example for a tDMRG
calculation with 49 bath spins is shown in Fig.~\ref{fig:49s}. The AHT results
are located between the tDMRG and the random noise results. Only
  a marginal difference between the different AHT curves is observed. The AHT result is
stable towards small fluctuations of the correlation function. So there is no
quantitative change in the AHT when the slightly time-dependent tDMRG autocorrelation is replaced by a completely constant one.
  As mentioned before, the improvement in the 2nd order of the AHT is minor, cf. Appendix~\ref{app:AHT}. The
plateau of the central spin autocorrelation still persists, as it only depends on
the variance for $t\rightarrow\infty$ and always emerges as long as
$\sigma^2\rightarrow\infty$ for $t\rightarrow\infty$. Note that there are no
corrections in 2nd order for a static autocorrelation
function. 

The plateau in the tDMRG result is located above the AHT
result. Furthermore, it is likely to decay on a longer time scale because we are
dealing with a finite system. The random noise simulation does not reach
the plateau of the AHT. 
  After the local minimum close to $t\approx 4$, the autocorrelation function of the central spin
$\braket{S^z(t)S^z(0)}$ increases but decays again shortly afterwards. The decay is slow but clearly visible.

 For a very large number of bath spins in
the limit $N\rightarrow\infty$, we expect that both the tDMRG and the random noise result
converge to the AHT result as the bath becomes more and more static.
  This is indeed the case, as shown in Fig.~\ref{fig:converge} for 49 to 999 bath
spins. The tDMRG curves converge from above towards the AHT result,
which is identical to the random noise simulation 
  for a constant autocorrelation function. The tDMRG result for $N=999$ bath
spins almost lies on top of the AHT result. The random noise simulation converges
from below. At the first sight, this seems to happen a little bit
  slower than for the tDMRG curves. However, one has to keep in mind that the
sampling in the simulation is always done for a finite number of fluctuations,
usually $M=100.000$. 
  Thus, there is always an error from the sampling of the order of $1/\sqrt{M}$ which is also visible on the
scale of the inset in Fig.~\ref{fig:converge}.
 
 \begin{figure}[tb]
  \centering
   \includegraphics[width=\columnwidth]{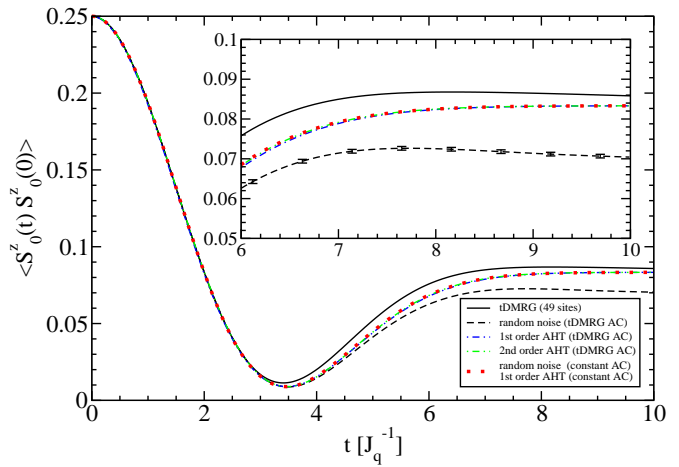}
    \caption{(Color online) Comparison between the tDMRG result for 49 bath spins, the
random noise simulation, and the AHT result in 1st and 2nd order. The AHT and the
random noise result have both been calculated with the tDMRG autocorrelation
function (tDMRG AC) and a constant autocorrelation function (constant AC) as input.
    The AHT result is fairly robust. The plateau of $1/12$ in the
autocorrelation of the central spin occurs in the random noise simulation only
when the bath fluctuations are completely frozen. There is no noticeable
difference between the AHT in 1st and 2nd order.}\label{fig:49s}
 \end{figure}	

 \begin{figure}[tb]
  \centering
   \includegraphics[width=\columnwidth]{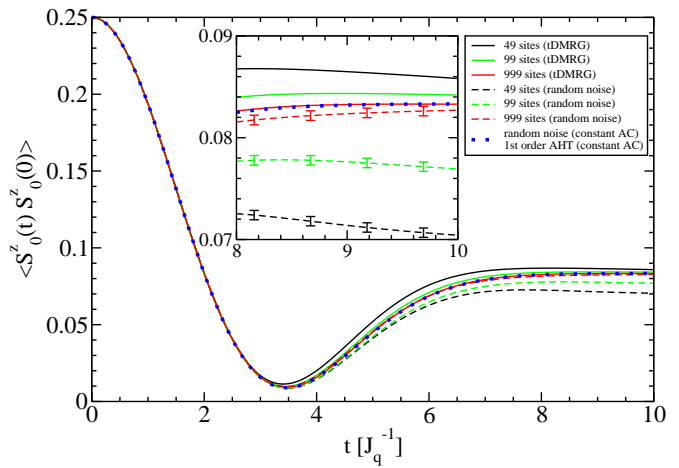}
    \caption{(Color online) Convergence of the tDMRG calculation and the random noise
simulation towards the AHT result for large system sizes. For the random noise
simulation, the autocorrelation functions obtained by tDMRG are used to
sample the fluctuations.}\label{fig:converge}
 \end{figure}	 

\section{Conclusion}

In this paper, we discussed the central spin model \eqref{eq:CSM} in the
low-field limit. The hyperfine coupling between central spin and bath is fully
captured by our numerical approach based on DMRG. 
We presented how the DMRG can be adopted for star-like spin clusters. In addition, we discussed
three different methods for the calculation of the real-time evolution: The
adaptive method based on the Trotter-Suzuki decomposition, 
time-step targeting with Krylov vectors, and the Chebychev expansion. The
calculation of observables at $T=\infty$ is realized via purified states.
Thereby, the exact trace can be calculated as expectation value of a single state at the cost
of doubling the size of the bath. 

For the real-time evolution, the best results are obtained with the Trotter-Suzuki decomposition. It combines
a slowly growing truncation error with moderate run-times and gives access to
relatively long times with good accuracy, even in 2nd order. The time-step targeting method
with Krylov vectors is applicable if smaller time scales are desired.
Then, it provides a higher accuracy than the Trotter-Suzuki decomposition because it
does not suffer from the Trotter-Suzuki error. But the truncation error grows
quite fast and much more CPU time is required compared to the Trotter-Suzuki
approach in 2nd order.

Furthermore, we demonstrated how the real-time evolution based on Chebychev
polynomials is implemented with DMRG. For the central spin model, this approach
displays a very fast growing truncation error. However, the situation in the
central spin model is special. Thus, we think that the Chebychev
expansion in the framework of DMRG may perform better for
other models, e.g., a Heisenberg chain. In a linear system, the entanglement created by any local
operator spreads slower than in the central spin model. Thus, the performance should increase.

In addition, we studied a semiclassical model where the bath is represented by a
classical fluctuating field. Therefore, we developed a systematic average
Hamiltonian theory based on the Magnus expansion. A comparison
of the tDMRG results, the AHT, and numerical simulations of the semiclassical
model revealed that the dynamic in the central spin model is well captured by
the semiclassical model for large bath sizes. In the final state of this paper, we became aware of a recent preprint by Witzel \textit{et al}.\cite{Witzel2013}
They also introduced a semiclassical approximation to describe a spin bath. In contrast to our study,
spin-flips between the central spin and the bath were neglected because the strong-field
limit was discussed. Thus the semiclassical model only comprised dephasing, no relaxation appears. 
Furthermore, the autocorrelation function of the semiclassical fluctuations
were obtained from a correlated cluster expansion.\cite{Yang2008a,Yang2009,Witzel2012} 

Future studies can include the behavior of the central spin model in a
magnetic field. Accordingly, the random fluctuations with a cylindric symmetry
could be discussed for the semiclassical model. 
 Nowadays, many numerical studies of pulses
and pulse sequences from dynamic decoupling are based on exact diagonalization techniques. 
In the future, these investigations
can be extended to systems containing a much larger number of bath spins, as the real-time evolution with DMRG based on the Trotter-Suzuki
decomposition is easily extendable to time-dependent Hamiltonians.

\acknowledgments

We would like to thank F. B. Anders, J. Hackmann, and J. Stolze for many useful discussions.
We gratefully acknowledge the financial support from the Studienstiftung
des deutschen Volkes (DS) and from the Mercator Stiftung (GSU).

\appendix

\section{4th order Trotter-Suzuki decomposition}\label{app:ST}

The 2nd order as well as the 4th order of the Trotter-Suzuki decomposition are
derived from the Magnus expansion \cite{Magnus1954,Blanes2009,Alverm2011} of the time-evolution
 operator. In this appendix, we derive the 4th order decomposition and the required expansion coefficients. 

For abbreviation, we denote a half sweep through the super block (e.g. from left
to right) as
\begin{align}
 P_{1N}\left(x\right)=e^{-\mathrm{i}xh_1} e^{-\mathrm{i}x h_2} \ldots e^{-\mathrm{i}x h_N}, \label{eq:P1N}
\end{align}
where $N$ is the number of bath spins. The local Hamiltonian $h_i$ contains the
complete interaction between bath spin $i$ and the central spin. The
corresponding back sweep to \eqref{eq:P1N} is denoted by $P_{N1}(x)$ and is simply
obtained by reversing the order of the local time-evolution operators in \eqref{eq:P1N}.

The 4th order of the Trotter-Suzuki decomposition is derived as the average Hamiltonian expression for a half sweep $P_{1N}$ or $P_{N1}$~\eqref{eq:P1N}, respectively. By the succesive application of
six half sweeps, one obtains 
\begin{align}
  \begin{split}
 &P_{1N}\left(\mu\right)
    P_{N1}\left(\lambda\right)
    P_{1N}\left(1\right) P_{N1}\left(1\right)P_{1N}\left(\lambda\right)
    P_{N1}\left(\mu\right) \\ &\quad\quad=e^{c_1+c_2+c_3+\mathcal{O}\left(\Delta
t^5\right)}.\label{eq:4ST_P} \end{split}
\end{align}
The operators $c_i$ in the exponential read
\begin{subequations}\label{eq:4ST_c}
\begin{align}
 c_1&= 2\Omega_1\left(1+\lambda+\mu\right) \\
c_2 &= \Omega_2\left(\lambda^2-\lambda^2+\mu^2-\mu^2\right) = 0\\
\begin{split}
c_3&=2\Omega_3\left(\lambda^3+\mu^3+1\right) \\
&+\left[\Omega_1,\Omega_2\right]\left(1+\mu^3+2\lambda\mu^2+2\mu^2-\lambda^3-2\lambda^2\right),\label{eq:4ST_c3}
\end{split}
\end{align}
\end{subequations}
where $\Omega_n$ denotes the $n$-th order contribution of the Magnus expansion.\cite{Alverm2011} The 1st order is given by the local decomposition of the Hamiltonian $H$
\begin{align}
 \Omega_1&=\sum\limits_{i=1}^N h_i.
\end{align}
The brackets in Eq.~\eqref{eq:4ST_c3} vanish for $\mu=-1/\sqrt[3]{2}=\lambda$ so that Eq.~\eqref{eq:4ST_P} corresponds to the desired time-evolution operator up to $\mathcal{O}(\Delta t^5)$. 

Now, we define 
\begin{align}
 \begin{split}
  e^{-\mathrm{i}H\Delta t }&= P_{1N}\left(\alpha\frac{\Delta t}{2}\right)\cdot
    P_{N1}\left(\beta\frac{\Delta t}{2}\right)\cdot
    P_{1N}\left(\gamma\frac{\Delta t}{2}\right) \\ &\quad\times
P_{N1}\left(\gamma\frac{\Delta
        t}{2}\right)\cdot P_{1N}\left(\beta\frac{\Delta t}{2}\right)\cdot
    P_{N1}\left(\alpha\frac{\Delta t}{2}\right)\\ &\quad+\mathcal{O}\left(\Delta
t^5\right). 
 \end{split}\label{eq:app_st}
\end{align}
for the 4th order of the Trotter-Suzuki decomposition.
The coefficients 
 \begin{subequations}
\begin{align}
  \alpha&=\frac{1}{2-2^{1/3}}=\beta \\ 
  \gamma&=-\frac{1}{2^{2/3}-1}
\end{align}
  \end{subequations}
are obtained by rescaling the solutions for $\lambda$ and $\mu$ from Eqns. \eqref{eq:4ST_c}.

If the 4th order decomposition is used to calculate the real-time evolution with
the adaptive method, three back and forth sweeps are required to proceed one
step $\Delta t$ in time. After $1/\Delta t$ applications, the Trotter-Suzuki error in 4th order accumulates to
$\mathcal{O}(\Delta t^4)$.

\section{Purified states}\label{app:Pur}

We consider a state $\ket{S_1,S_2}$ of two half-integer spins, e.g., two nearest neighbors taken from a chain. The operators acting on site 1 and 2 are denoted by $\vec{S}_{i,\mathrm{r}}$. A generic Hamiltonian
for the two spins is $H_\mathrm{r}=\vec{S}_{1,\mathrm{r}}\cdot\vec{S}_{2,\mathrm{r}}$. In addition, an auxiliary spin with operator $\vec{S}_{i,\mathrm{a}}$ is introduced to each real spin. 

The action of $H_\mathrm{r}$ on the initial state $\ket{S_1,S_2}$ is given as 
\begin{subequations}
 \begin{align}
    H_\mathrm{r}\ket{S_1,S_2}&=\vec{S}_{1,\mathrm{r}}\cdot\vec{S}_{2,\mathrm{r}}\ket{S_1,S_2} \\ 
    &=-\vec{S}_{1,\mathrm{a}}\cdot\vec{S}_{2,\mathrm{r}}\ket{S_1,S_2},
 \end{align}
where the spin $S_1$ has been swapped with its auxiliary antiparallel spin sitting on the same site. Swapping the 2nd spin with its auxiliary one cancels the minus sign again so that
the action of $H_\mathrm{r}$ on the initial state is given as
\begin{align}
     H_\mathrm{r}\ket{S_1,S_2}&=\vec{S}_{1,\mathrm{a}}\cdot\vec{S}_{2,\mathrm{a}}\ket{S_1,S_2}.
\end{align}
\end{subequations}
The latter expression implies
\begin{align}
 H_\mathrm{r}\ket{S_1,S_2}&= H_\mathrm{a}\ket{S_1,S_2},
\end{align}
where the Hamiltonian $H_\mathrm{a}=\vec{S}_{1,\mathrm{a}}\cdot\vec{S}_{2,\mathrm{a}}$ acts on the auxiliary sites only. Hence, the action of
both Hamiltonians on the initial state $\ket{S_1,S_2}$ is the same. 

Consequently, the action of the Hamiltonian
\begin{align}
 H&=H_\mathrm{r}-H_\mathrm{a}
\end{align}
on a purified inital state $\ket{0}$ is always zero because all contributions compensate each other. For the application of the time-evolution operator $U=e^{-\mathrm{i}Ht}$, this implies
\begin{align}
 e^{-\mathrm{i}\left(H_\mathrm{r}-H_\mathrm{a}\right)}\ket{0}&=\ket{0}, \label{eq:APP_PUR_U}
\end{align}
which is valid as long as all real sites are prepared as singlets (or $m=0$ triplets) with their corresponding auxiliary sites. The property \eqref{eq:APP_PUR_U} is destroyed if any operator is applied to the real sites
so that $\ket{0}$ is changed to a state different from a product of singlets. To our knowledge, this analytic argument
has not been present in the literature so far.

Note, however, that the advantage of using $H=H_\mathrm{r} - H_\mathrm{a}$ depends on the topology.
In a chain with nearest-neighbor interactions, the benefit is largest because a local perturbation at site $j$
will be felt at site $j+n$ only at $n$ applications of $H$. In the star topology of the central spin model
the situation is different. Applying an operator to the central spin and subsequently $H$ destroys the singlet
character already at every bath site.

\section{2nd order AHT}\label{app:AHT}

Before we introduce the 2nd order of the AHT, we consider the unaveraged 1st
order expression from Eq.~\eqref{eq:AHT_unavg} prior to averaging. It is rewritten in the form
\begin{align}
 f\left(t\right)&=\frac{1}{12}+\frac{1}{6}X\left(1\right) \label{eq:2ndAHT_f}
\end{align}
with 
\begin{align}
 X\left(a\right)&=4\pi\int_0^\infty \mathrm{d}v\, v^2 \cos\left(va\right)
P\left(v\right). \label{eq:2ndAHT_X}
\end{align}
The probability distribution $P\left(v\right)$ is spherical according to the
symmetry of the problem, i.e., it depends on the modulus $v=|\vec{v}|$ only. The previous expression for $X$ is generalized by
replacing $P(v)$ with 
the probability distribution of one compoment of $\vec{v}$, e.g. the
$z$-compoment, via
\begin{align}\begin{split}
 p\left(v_z\right)&=\iiint  \mathrm{d}^3v\, P\left(v\right)\delta\left(v_z-v\cos\theta\right) \\ 
  &=2\pi\int_{|v_z|}^\infty \mathrm{d}v\, v P\left(v\right).\end{split}
\end{align}
This equation is differentiated with respect to $v_z$ and inserted into Eq.
\eqref{eq:2ndAHT_X}. Hence, one obtains
\begin{align}
  X\left(a\right)&=-2\int_0^\infty \mathrm{d}v\, v \cos\left(va\right) p'\left(v\right).
\end{align}
By partially integrating this expression with respect to $a$, it is reduced to
\begin{align}
 X\left(a\right)&=\tilde{p}\left(a\right)+a\tilde{p}'\left(a\right)
\label{eq:2ndAHT_X_p}
\end{align}
where the Fourier transform $\tilde{p}(a)=\int_{-\infty}^\infty \mathrm{d}v\, p(v)
e^{\mathrm{i}va}$ of $p(v)$ has been introduced.

Thereby, we obtained a general expression for the correlation function
\eqref{eq:AHT_unavg}. Just the Fourier transform $\tilde{p}(a)$ has to be
calculated, which is
nothing else but the mean value of $e^{\mathrm{i}av}$. In the following, we show how
this expression is calculated and evaluated.

The Magnus expansion\cite{Magnus1954,Blanes2009} up to 2nd order reads
\begin{align}\begin{split}
 U\left(t\right)&=\exp\left\{-\mathrm{i}\int_0^t \mathrm{d}t'\,H\left(t'\right)\right. \\
  &\quad\left.+\frac{1}{2}\int_0^t \mathrm{d}t_1\int_0^{t_1} \mathrm{d}t_2\,
\left[H\left(t_1\right),H\left(t_2\right)\right]\right\}.
\end{split}
\end{align}
It can always be written in the form 
\begin{align}
  U\left(t\right)&=e^{-\mathrm{i}\vec{v}\cdot\vec{S}}
\end{align}
with
\begin{align}\begin{split}
 v_x&=\int_0^t \mathrm{d}t'\,\eta_x\left(t'\right) \\ 
&\quad+\frac{1}{2}\int_0^t \mathrm{d}t_1\int_0^t \mathrm{d}t_2\,
\eta_y\left(t_1\right)\eta_z\left(t_2\right)\text{sgn}\left(t_1-t_2\right)\end{split}
\end{align}
and cyclic in the components $x,y$ and $z$. This expression is the desired argument of the exponential in $\tilde{p}(a)$, just the average 
\begin{align}
 \tilde{p}\left(a\right)&=\overline{e^{\mathrm{i}av_x}} \label{eq:2ndAHT_p}
\end{align}
with respect to all three components $\eta_x$, $\eta_y$, and $\eta_z$ remains to be calculated. 

The mean value for
Gaussian fluctuations can be evaluated analytically by applying the general identity
\begin{align}
 \begin{split}
  &\overline{\exp\left[\int_{t_1}^{t_2}\, \mathrm{d}\tau\,
a\left(\tau\right)\eta\left(\tau\right)\right]}\\
    &\quad=\exp\left[\frac{1}{2}\iint_{t_1}^{t_2}
\mathrm{d}t\,\mathrm{d}\tau\,a\left(t\right)g\left(t-\tau\right)a\left(\tau\right)\right].
 \end{split}
\end{align}
In the contribution resulting from the 1st order of the Magnus expansion, only the mean value for $\eta_x$ has to be
calculated. Therefore, one obtains
\begin{align}
 \overline{\exp\left[\mathrm{i}a\int_0^t \mathrm{d}t'\,
\eta_x\left(t'\right)\right]}^x&=\exp\left[-a^2G\left(t\right)\right] 
\end{align}
where 
\begin{align}
 G\left(t\right)&=\int_0^t \mathrm{d}t_1\int_0^{t_1} \mathrm{d}t_2\, g\left(t_1-t_2\right)
\end{align}
is even.

For the contribution from the 2nd order of the Magnus expansion, the average with respect to $\eta_y$ is carried
out analytically
\begin{align}\begin{split}
\overline{ \exp\left[\frac{\mathrm{i}a}{2}\int_0^t \mathrm{d}t_1\int_0^t \mathrm{d}t_2\,
\eta_y\left(t_1\right)\eta_z\left(t_2\right)\text{sgn}\left(t_1-t_2\right)\right
]}^y \\
\quad= \exp\left[-\frac{a^2}{8}\int_0^t \mathrm{d}t_1^{\phantom{'}}\int_0^t \mathrm{d}t_1'\,
\alpha\left(t_1^{\phantom{'}}\right)g\left(t_1^{\phantom{'}}-t_1'\right)\alpha\left(t_1'\right)\right]\end{split}\label{eq:APP_avg_y}
\end{align}
where
\begin{align}
 \alpha\left(t_1\right)&:=\int_0^t
\mathrm{d}t_2\,\text{sgn}\left(t_1-t_2\right)\eta_z\left(t_2\right)
\end{align}
still depends on $\eta_z$. After rearranging the integrals in the latter expression, the
intermediate result for $\tilde{p}(a)$ from Eq. \eqref{eq:2ndAHT_p} can be written as
\begin{subequations}
\begin{align}\begin{split}
\tilde{p}\left(a\right)&=e^{-a^2G\left(t\right)} \\ 
&\quad\times\overline{e^{-\frac{a^2}{8}\int_0^t\mathrm{d}t_2^{\phantom{'}}\int_0^t\mathrm{d}t_2'\,
\eta_z\left(t_2'\right)A\left(t_2',t_2^{\phantom{'}}\right)\eta_z\left(t_2^{\phantom{'}}\right)}}^z.\label{eq:2ndAHT_intermediate}
\end{split}\end{align}
The two integrations with respect to $t_1^{\phantom{'}}$ and $t_1'$ in $A(t_2',t_2^{\phantom{'}})$ are carried out analytically and one obtains 
\begin{align}
 \begin{split}   
A\left(t_2',t_2^{\phantom{'}}\right)&:=\iint_0^t\mathrm{d}t_1\,\mathrm{d}t_1'\,\text{sgn}
\left(t_1'-t_2'\right)g\left(t_1'-t_1^{\phantom{'}}\right)\\ 
&\quad\quad\times\text{sgn}\left(t_1^{\phantom{'}}-t_2\right) \\
  &=2\left[G\left(t_2-t\right)+G\left(t_1-t\right)-2G\left(t_2-t_1\right)\right.\\
  &\quad\quad\left.+G\left(t_1\right)+G\left(t_2\right)-G\left(t\right)\right].
 \end{split}
\end{align}
\end{subequations}

\begin{figure}[bt]
  \centering
   \includegraphics[width=\columnwidth]{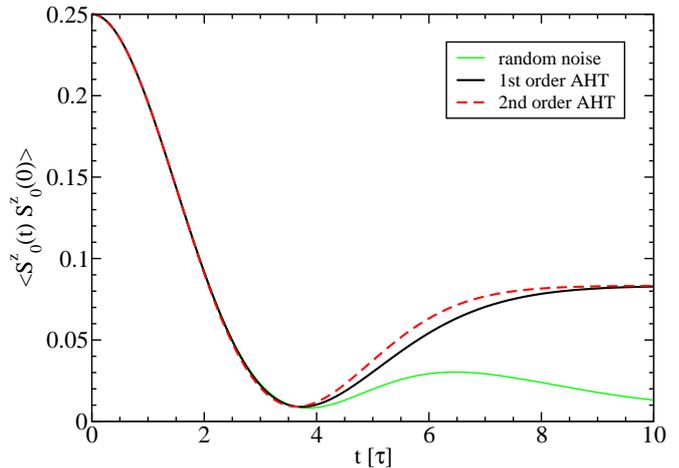}
    \caption{(Color online) AHT in 1st and 2nd order for the exemplary correlation function $g(t)=1/4\,e^{-|t|/(8\tau)}$.}\label{fig:APP_AHT}
 \end{figure}

The average with respect to $\eta_z$ still remains and cannot be calculated in a
closed form analytically, even though it is a Gaussian average. Thus, we choose a simple numerical approach based on the
discretization of time $t$ in
$N$ intervals of width $\Delta t=t/N$. 
Then, the average of $\vec{\eta}_z=(\eta_z(t_1),\ldots,\eta_z(t_N))^\top$ is
carried out with respect to an $N$-dimensional probability distribution and the
integrals in the exponential are replaced by sums. Consequently, the Fourier
transform \eqref{eq:2ndAHT_intermediate}
reads
\begin{align}
 \begin{split}
\tilde{p}\left(a\right)&=\frac{e^{-a^2G\left(t\right)}}{\sqrt{\text{det}\,\mathbf{
M}}} \int_{-\infty}^\infty \frac{\mathrm{d}^N\eta_z}{\left(2\pi\right)^{N/2}}\,
e^{-\frac{1}{2}\vec{\eta}_z^\top\mathbf{M}^{-1}\vec{\eta}_z^{\phantom{\top}}}
e^{-\frac{1}{2}\vec{\eta}_z^\top\mathbf{P}\vec{\eta}^{\phantom{\top}}_z}.\label{eq:2ndAHT_P}
\end{split}
\end{align}
\begin{subequations}
The matrix $\mathbf{M}$ is the covariance matrix defined by 
\begin{align}
M_{ij}&:=g\left(t_j-t_i\right),
\end{align}
while
\begin{align}
 P_{ij}&:=\frac{a^2}{4} A\left(t_j,t_i\right) \Delta t^2
\end{align}
contains the correction of the 2nd order.
The discretized time steps with width $\Delta t=t/N$ are given as
\begin{align}
 t_i&=\left(i-\frac{1}{2}\right)\frac{t}{N}, \ \ \ i\in\{1,\ldots,N\}.
\end{align}
 \end{subequations}
Note that the structure of the correction in Eq. \eqref{eq:2ndAHT_P} is also bilinear in $\eta_z$. Hence, the
$N$-dimensional integration is carried out easily and the final result for the
Fourier transform
reads
\begin{align}
 \tilde{p}\left(a\right)&=\frac{1}{\sqrt{\text{det}\left(\mathds{1}+\mathbf{PM}
\right)}} e^{-a^2G\left(t\right)}. \label{eq:2ndAHT_p_final}
\end{align}
In total, the 2nd order of the AHT has led to a renormalization of the probability
distribution by a factor $1/\sqrt{\mathrm{det}\,(\mathds{1}+\mathbf{PM})}$.  
With the Fourier transform and Eqns.~\eqref{eq:2ndAHT_f} and \eqref{eq:2ndAHT_X_p}, one has obtained
the final expression for the autocorrelation function $f(t)$.

Fortunately, only a rough discretization of $t$ is sufficient. In practice,
we use 20-30 time intervals for evaluating~\eqref{eq:2ndAHT_f} up to
$t=10$. As an example, the effect of the 2nd order AHT is illustrated for the correlation function
$g(t)=1/4\, e^{-|t|/(8\tau)}$ in Fig.~\ref{fig:APP_AHT}. The 2nd order leads to a faster stabilization of the plateau
in the autocorrelation function of the central spin. Note that the plateau does not exist in the numerical simulation. There,
the autocorrelation function of the central spin decays completely for $t\gg0$. This must be attributed to effects beyond the Magnus
expansion because the plateau is not altered by the 2nd order Magnus corrections.

\end{document}